\documentclass[aps,prd,twocolumn,groupedaddress]{revtex4}
\usepackage{epsf,epsfig,graphics,graphicx}
\begin{document}
\title{Running spectral index and mode-mode correlation of inflationary perturbations from off-equilibrium effects }
\author{Da-Shin Lee}
 \affiliation{Department of Physics, National Dong-Hwa
University, Hua-lien, Taiwan 974, R. O. C.}
\author{Li-Zhi Fang}
 \affiliation{Department of Physics, University of Arizona,
         Tucson, AZ 85721, U. S. A.}
\author{Wolung Lee}
\author{Yeo-Yie Charng}
 \affiliation{Institute of
Physics, Academia Sinica, Taipei, Taiwan 115, R. O. C.}

\date{\today}

\begin{abstract}

We study  the  off-equilibrium effects of inflaton on the dynamics
of  primordial perturbations in the $O(N)$ model. A
self-consistent off-equilibrium formalism is employed to
investigate the evolution of the inflationary background field and
its fluctuations with the back-reaction effects. We find two
observable remains left behind the off-equilibrium processes: the
running spectral index of primordial density perturbations and the
correlations between perturbation modes in phase space, which
would serve as  the imprints  to probe the epoch of inflation,
even beyond.

\end{abstract}

\maketitle

In the inflationary scenario, the primordial perturbations of the
universe originate from vacuum fluctuations of the scalar
field(s), the inflaton $\phi$, driving the inflation. If the
dynamics of the fluctuations is approximated by single massless
free field during the inflation, the power spectrum of curvature
perturbations for the mode $k$, ${\cal P}_{\cal R}(k)$ is roughly
$k$-independent~\cite{lid}. However the $k$-dependence, or the
running of the spectral index, revealed by the recently released
Wilkinson Microwave Anisotropy Probe (WMAP) data can be as large
as $dn(k)/d\ln k=(d/d\ln k)^2 \ln{\cal P}_{\cal R}(k)\simeq
-0.055{\ \rm to \ }-0.077$~\cite{pei}. Hence, the origin of the
primordial perturbations cannot be solely accommodated with the
quantum fluctuations of  single free field. In a slow-roll
inflation, one has $d/d\ln k =(1/H)d/dt = -(1/8\pi
G)(V'(\phi)/V(\phi))(d/d\phi)$, and thus the derivative of the
power spectrum $(d/d\ln k)^n\ln{\cal P}_{\cal R}(k)$ will no
longer be negligible for the second order ($n=2$) as long as the
third or higher derivatives of the inflaton potential becomes
substantial. To fit in with the running of the spectral index,
models beyond single scalar field with quadratic potential
$V(\phi)$ have been proposed accordingly~\cite{kaw}. In this
context, the running index of the perturbation power spectrum is
considered as an essential feature due to the interactions or the
self-interactions of the inflaton(s).

It should be pointed out here that in deriving the derivative of
the power spectrum above, the effective potential $V(\phi)$ is
involved. However, this approach may be problematic, when in
particular, the one-loop effective potential becomes complex where
the background field $\phi$ is in the region of
$V''(\phi)<0$~\cite{do}. The imaginary part of the effective
potential would inevitably lead to a dynamically unstable
state~\cite{we}. This so-called ``spinodal instability'' will
allow long wavelength fluctuations to grow non-perturbatively.
Therefore, the primordial perturbations must be dealt consistently
with a different method to account for the amplification of vacuum
fluctuations in the presence of spinodal instabilities. This
motivates us to study the self-interaction effects of the inflaton
on the dynamics of primordial perturbations within a context of
the self-consistent off-equilibrium formalism.  We will use the
$O(N)$ model of inflation as an example. The $O(N)$ model with
spontaneously broken symmetry  has been extensively used in
modelling the quantum off-equilibrium processes in the early
universe, as well as the chiral phase transition in relativistic
heavy ion collision~\cite{boyn1,boyn2,ra,baa}. Here, we will focus
on searching for the possible observable imprints caused by the
off-equilibrium evolution of inflationary primordial
perturbations~\cite{lee1}.

Consider the dynamics of an inflation driven by a field ${\bf\Phi}$
of the $O(N)$ vector model with spontaneous symmetry breaking. The
action is defined by
\begin{equation}
{\mathcal S}=\int d^4x {\mathcal L}= \int d^4x\sqrt{-g}\left [
\frac{1}{2}g^{\mu\nu}
     \partial_{\mu} {\bf \Phi} \cdot \partial_{\nu}{\bf \Phi} -
        V({\bf \Phi}\cdot{\bf \Phi})\right ],
\end{equation}
where $V({\bf \Phi}\cdot{\bf \Phi})$ is a self-interaction
potential given by
\begin{equation}
V({\bf \Phi}\cdot{\bf \Phi}) = \frac{\lambda}{8N} \left ({\bf
\Phi}\cdot{\bf \Phi} - \frac{2Nm^2}{\lambda}  \right )^2 \, .
\label{pot}
\end{equation}
As an inflaton, the $N$ components of the field generally are
represented as ${\bf \Phi} = (\sigma, \vec{\pi})$, where $\vec{
\pi} $ represent  $N-1$ scalar fields. The cosmic inflation is
characterized by the state in which the component $\sigma$ has a
spatially homogeneous expectation, i.e. $\sigma$ can be decomposed
into a background field plus fluctuations around the background
field as
\begin{equation}
 \sigma({\bf x},t) =\sqrt{N}\phi(t) + \chi({\bf x},t) \,
\end{equation}
with the expectation value  $\langle\sigma({\bf x},t)\rangle=
\sqrt{N}\phi(t)$ and thus,  $\langle \chi ({\bf x},t) \rangle =0$.
During the inflationary epoch, the background space-time can be
described by a spatially flat Friedmann-Robertson-Walker metric,
\begin{equation}
ds^2=  g_{\mu\nu} dx^{\mu} dx^{\nu}= dt^2 -
a^2(t)\delta_{ij}dx^idx^j,
\end{equation}
where the scale factor $a=\exp({Ht})$ with the expansion rate
$H=\sqrt{8\pi G\rho/3}$ determined by the mean of energy density
of the inflaton field, $\rho$.

 In order to take account of the growth of fluctuations due to spinodal
 instabilities as we will see later,
 we will
employ the method of the Hartree factorization, which approximates
the potential $V$ with an effective quadratic potential while
keeping $N$ finite~\cite{boyn2,baa}. Then the equation of motion
of the background field $\phi$ can be directly obtained from the
Hartree-factorized Lagrangian by means of the tadpole condition
$\langle \chi ({\bf x},t) \rangle =0$ given by
\begin{equation}
\ddot{\phi} (t)+3H\dot{\phi} (t)+\left[M_{\chi}^2(t)-\lambda \phi^2 (t)
\right]\phi (t)=0 \, . \label{classfieldeq}
\end{equation}
We then decompose $\chi(t)$ and $\vec{\pi}(t)$ in the Fourier
basis. In the Heisenberg picture, one has
\begin{eqnarray}
\chi({\bf x}, t)& = &  \int\frac{d^3k}{8\pi^3} \, \left[ b_{\bf
k}f_{\chi, \bf k} (t)  + b^\dagger_{\bf -k}f^{\star}_{\chi, \bf
-k}(t) \right ] e^{i {\bf
k}\cdot {\bf x}},  \nonumber \\
 \pi_{i}({\bf x}, t)& = &   \int\frac{d^3k}{8\pi^3} \,  \left[ a_{i \bf k}f_{\pi, \bf k} (t) +
a^\dagger_{i \bf -k}f^{\star}_{\pi, \bf -k}(t)  \right ] e^{i {\bf
k}\cdot {\bf x}}, \nonumber \\
 \label{chipi}
\end{eqnarray}
where $a_{i \bf k}$, $b_{ \bf k}$ and $a^\dagger_{i \bf k}$,
$b^\dagger_{ \bf k}$ are the creation and annihilation operators
which obey the commutation relations. The equations of the mode
functions $f_{\chi, \bf k} (t)$ and $f_{\pi, \bf k} (t)$ can be
found from the Heisenberg field equations as follows:
\begin{eqnarray}
&& \left[\frac{d^2}{dt^2}+3H\frac{d}{dt}+
\frac{k^2}{a^2}+M_{\chi}^2 (t) \right ]f_{\chi, \bf k}(t)=0, \nonumber \\
&& \left[\frac{d^2}{dt^2}+3H\frac{d}{dt}+
\frac{k^2}{a^2}+M_{\pi}^2 (t) \right ]f_{\pi, \bf k}(t)=0.
\label{modeeq}
\end{eqnarray}
The time dependent mass terms are give by
\begin{eqnarray}
M^2_{\chi}(t) &=& -m^2 +  \frac{3\lambda}{2}\phi^2 (t) +
\frac{3\lambda}{2N}\langle\chi^2\rangle (t) \nonumber \\
&+& \frac{\lambda}{2}\left(1-\frac{1}{N}\right)\langle
\psi^2\rangle (t) +\xi R \, ,
\nonumber \\
 M^2_{\pi} (t) &=& -m^2 +  \frac{\lambda}{2}\phi^2(t) +
\frac{\lambda}{2N}\langle\chi^2\rangle (t) \nonumber \\
&+& \frac{\lambda}{2}\left(1+\frac{1}{N}\right)\langle
\psi^2\rangle (t) +\xi R \, , \label{selfconmass}
\end{eqnarray}
where $\langle \psi^2\rangle$ is defined by
$\langle\vec{\pi}^2\rangle=(N-1)\langle \psi^2\rangle$. The
initial conditions of the mode functions can be specified as
effective free massive scalar fields in an expanding
universe~\cite{boyn2}:
\begin{eqnarray}
f_{\chi, \bf k} (0) &=& \frac{1}{\sqrt{ 2 [k^2 +
M^{2}_{\chi}(0)]}},\dot{f}_{\chi, \bf k} (0) =- i
 \sqrt{\frac{k^2 + M^{2}_{\chi}(0)}{ 2 } } \,  ; \nonumber \\
 f_{\pi, \bf k} (0) &=& \frac{1}{ \sqrt{ 2 [k^2 + M^{2}_{\pi}(0)]}}
 ,   \dot{f}_{\pi, \bf k} (0) =- i
 \sqrt{\frac{k^2 + M^{2}_{\pi}(0)}{ 2 } } \, , \nonumber \label{initial}
\end{eqnarray}
where we have set $ a(0)=0 $. The values of $M^{2}_{\chi}(0)$ and
$M^{2}_{\pi}(0)$ depends on the details of the onset of the
inflation.

Finally, to close these equations self-consistently, the terms
$\langle \chi^2\rangle (t)$ and $\langle \psi^2\rangle (t)$ in the
mass-squared [Eq.(\ref {selfconmass})] can be determined by the
mode functions which involve the divergences associated with the
loop integrals. To get the self-consistent renormalized quantities
defined as
\begin{eqnarray}
\langle \chi^2\rangle_R (t) &=& \int^{\Lambda} \frac{d^3k}{8\pi^3}
|f_{\chi, {\bf k}}(t)|^2 -
 \frac{1}{8\pi^2} \frac{\Lambda^2}{a^2} \, , \nonumber \\
\langle \psi^2\rangle_R (t) &=& \int^{\Lambda} \frac{d^3k}{8\pi^3}
|f_{\pi, {\bf k}}(t)|^2 -
 \frac{1}{8\pi^2} \frac{\Lambda^2}{a^2} \, ,
  \label{selfconeqrenorm}
\end{eqnarray}
one has to absorb the divergences into renormalization of  the
bare mass and the bare coupling constant. However, for a weak
coupling $\lambda < 10^{-14}$ in a typical inflation model, the
logarithmic divergences can be neglected.
\begin{figure}[t]
\begin{center}
\leavevmode \epsfxsize=2.0in \epsffile{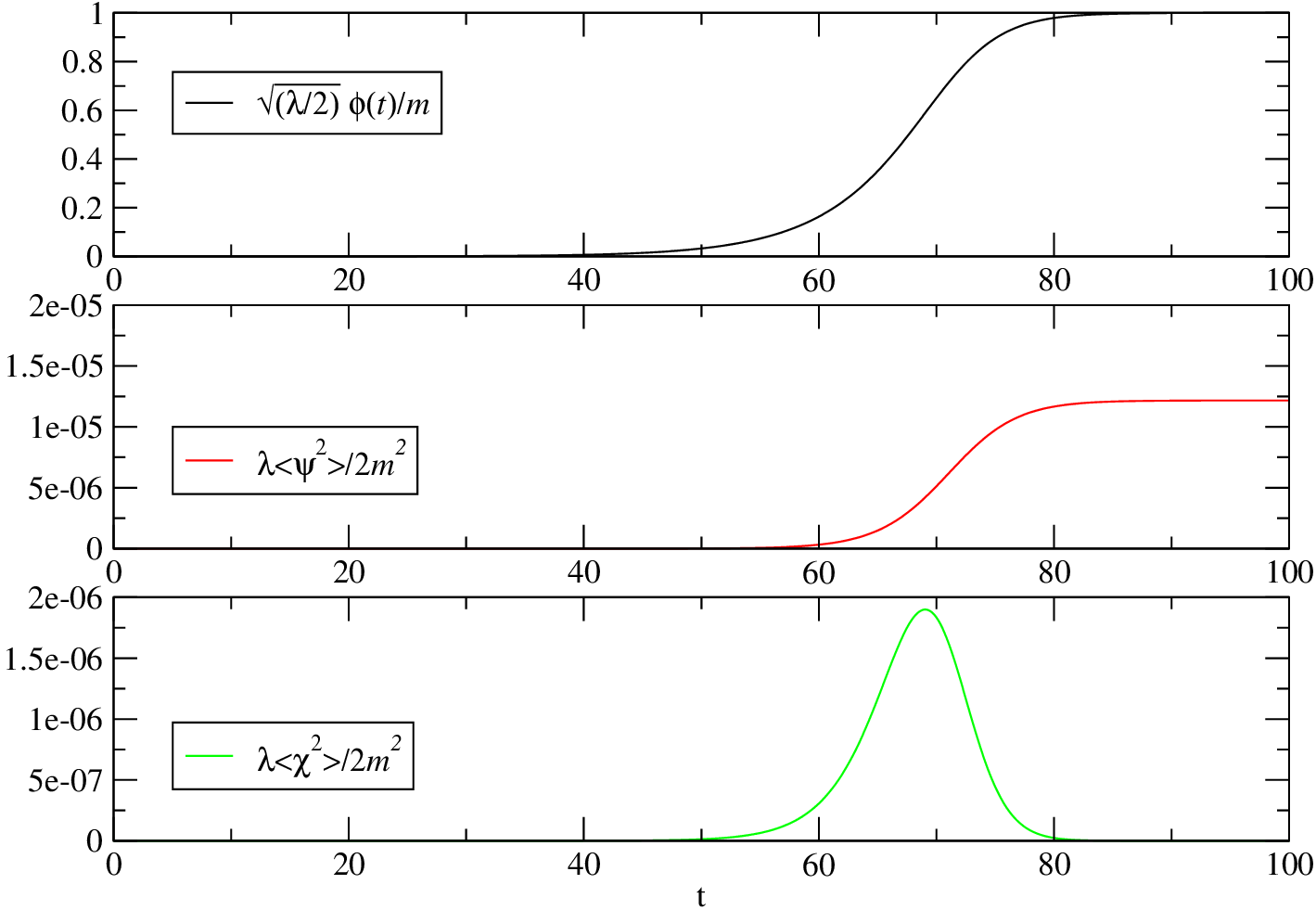}
\end{center}
\caption{The evolution of $\sqrt{\frac{\lambda}{2} }
\frac{\phi}{m}$, $ \frac{\lambda}{2 m^2} \langle \psi^2 \rangle$,
and $ \frac{\lambda}{2 m^2} \langle \chi^2 \rangle$ respectively
vs. $ t  $ ( in units of $m^{-1}$ ) for initial conditions
$\phi(0) \approx 0$, $\dot{\phi}(0)=0$,and the parameters
$h=\frac{H}{m} =2.0, \lambda =10^{-14}$ in the case of $N=4$.
 } \label{fig1}
\end{figure}

Figure 1 plots a self-consistent solution of the inflationary
background field $\phi (t) $ and the fluctuations $\langle \psi^2
\rangle (t)$, $\langle \chi^2 \rangle (t) $, in which the
parameters are taken to be  $ h=H/m =2.0$, $\lambda = 10^{-14}$,
$N=4$, and $M^{2}_{\chi}(0)= M^{2}_{\pi}(0)\simeq m^2 > 0$. In
fact, the value we choose for $ H/m $ above can be determined by
the energy density of the inflaton field. During the inflation,
the energy density of the inflaton is dominated by that of the
potential energy that leads to $ (H/m)^2 \approx N m^2/M^2_{\rm
Pl} \lambda \approx {\cal O} (1)$ for $ N \approx {\cal O} (1)$,
$\lambda \approx 10^{-14}$,  and the inflaton mass $m \approx
10^{12}$ Gev. We see that $\phi(t)\simeq 0$ and $
\langle\psi^2\rangle(t)\simeq\langle\chi^2\rangle(t)\simeq 0$
during the first stage of the inflation as $t<40m^{-1}$ in the
spinodal regime where the initially positive $M^{2}_{\chi}$ and
$M^{2}_{\pi}$ turn negative. The spinodal instabilities become
important eventually, and leads to the significant growth of both
$\langle\psi^2\rangle(t)$ and $\langle\chi^2\rangle(t)$ starting
at $t= 50m^{-1}$. Then, the spinodal condition is only weakly
satisfied over a span of $70m^{-1}\leq t \leq 80m^{-1}$ where the
background $\phi$ starts the rapid falling into the valley of the
inflaton potential. Finally, the process of spinodal instability
terminates at time $t_e$ given by $
 M^2_{\pi} (t_e) \approx 0 $.
  It renders $t_e \simeq (70-80) m^{-1}$. The background inflaton
$\phi $ sets in the minimum of a broken symmetry phase at the time
$ t_{\rm e}$ where $\phi\simeq\sqrt{2/\lambda}m$, which signifies
the end of inflation. Therefore, the number of total inflationary
e-foldings is $N_{\rm e} \approx Ht=2m t \approx 160$.  It is
evident that most of energy density of the background $\phi$ field
is transferred to the fluctuations $\langle\psi^2\rangle$ via the
production of $\pi$ modes during the inflation. Since $M^2_{\pi}
(t_e)\simeq 0$, these $\pi$ modes are in fact the massless
Goldstone modes of the broken symmetry phase. The Goldstone
theorem is fulfilled dynamically. Although the Hubble parameter
$H$ is treated here as a constant, the energy density of the
inflaton evolves dynamically.  In fact, the time dependent $H$ can
be determined by the evolution of the inflaton field $ \phi (t)$
as the effects of quantum fluctuations ($ \approx 10^{-5}$ ) on
the dynamics of the Hubble parameter is negligible. The
approximation of the constant $H$ can be argued to beak down  when
the $\dot\phi (t)$ reaches the maximum value at $ t=70m^{-1}$. A
truly self-consistent approach has to involve the correct dynamics
of the Hubble parameter through the semi-classical Einstein
equations as in Ref.~\cite{boyn2}. However, the qualitative
features of the above solutions are shown to remain the same.

Using the numerical solutions, we compute the power spectrum of
primordial perturbations.  The power spectrum of primordial
perturbations is described by ${\cal P}(k)=\langle |\delta_{\bf
k}|^2 \rangle$ where the mass density perturbations are determined
by the gauge invariant quantity~\cite{ko}
\begin{equation}
\delta_{\bf k} = \left .\frac{\delta \rho}{\rho+p} \right |_{k=aH}
\, . \label{masspert}
\end{equation}
The mass density fluctuation $\delta \rho$ originates from the
field fluctuations $\pi_{i}$ and $\chi$  given by
\begin{eqnarray}
\frac{\delta \rho}{N} &=& \left ( 1- \frac{1}{N}\right ) \left
[\frac {1}{2}\langle \dot{\psi}^2\rangle
   + \frac{1}{2a^2}\langle (\nabla \psi)^2\rangle
   -\frac{1}{2}m^2\langle \psi^2\rangle \right. \nonumber \\
   & +& \left. \frac{\lambda}{4}\phi^2\langle \psi^2\rangle
   +\frac{\lambda}{4N}\langle \chi^2\rangle\langle \psi^2\rangle\right.
   +\left. \frac{\lambda}{8}\left(1+\frac{1}{N}\right) \langle
\psi^2\rangle^2 \right] \nonumber \\
&+& \frac{1}{N}\left [\frac{1}{2}\langle\dot{\chi}^2\rangle
 + \frac{1}{2a^2}\langle (\nabla \chi)^2\rangle
   - \frac{1}{2}m^2\langle \chi^2\rangle
   +  \frac{3\lambda}{4}\phi^2\langle \chi^2\rangle \right.
   \nonumber \\
  &+& \left.\frac{\lambda}{4}\left(1-\frac{1}{N}\right)\langle\psi^2\rangle\langle\chi^2\rangle
  + \frac{3\lambda}{8N}\langle \chi^2\rangle^2 \right ] \, .
\label{deltadensity}
\end{eqnarray}
The term summing up the energy density and the pressure, $\rho+p$,
in Eq. (\ref{masspert}) can be obtained by
\begin{eqnarray}
 \frac{\rho+p}{N}&=&\dot{\phi}^2+
\left ( 1- \frac{1}{N}\right ) \left[\langle\dot{\psi}^2\rangle +
 \frac{1}{a^2}\langle(\nabla\psi)^2\rangle \right] \nonumber \\
 &+&
\frac{1}{N}\left[\langle\dot{\chi}^2\rangle +
 \frac{1}{a^2}\langle(\nabla\chi)^2\rangle \right ] \, .\label{pressure}
\end{eqnarray}
The spectral index $n(k)-1 = dP(k)/d\ln k$ and its $k$-dependence
$dn(k)/d\ln k$ are computed numerically as shown in Fig.2. The
value of $n(k)$ varies from unity at the larger scales as $t <
60m^{-1}$ to about $n=1+4(\nu-3/2)\simeq 1.32$ at the smaller
scale as $t>80m^{-1}$. Apparently, there is a significant
index-running $dn(k)/d\ln k \simeq 0.015$ in the wavelength range
corresponding to the horizon-crossing times during $60< mt <80$.
The running of the spectral index is in fact due to the energy
transfer from the inflationary background field to the
fluctuations which certainly can not be obtained from the
classical effective potential approach.

\begin{figure}[t]
\begin{center}
\leavevmode \epsfxsize=2.0 in \epsffile{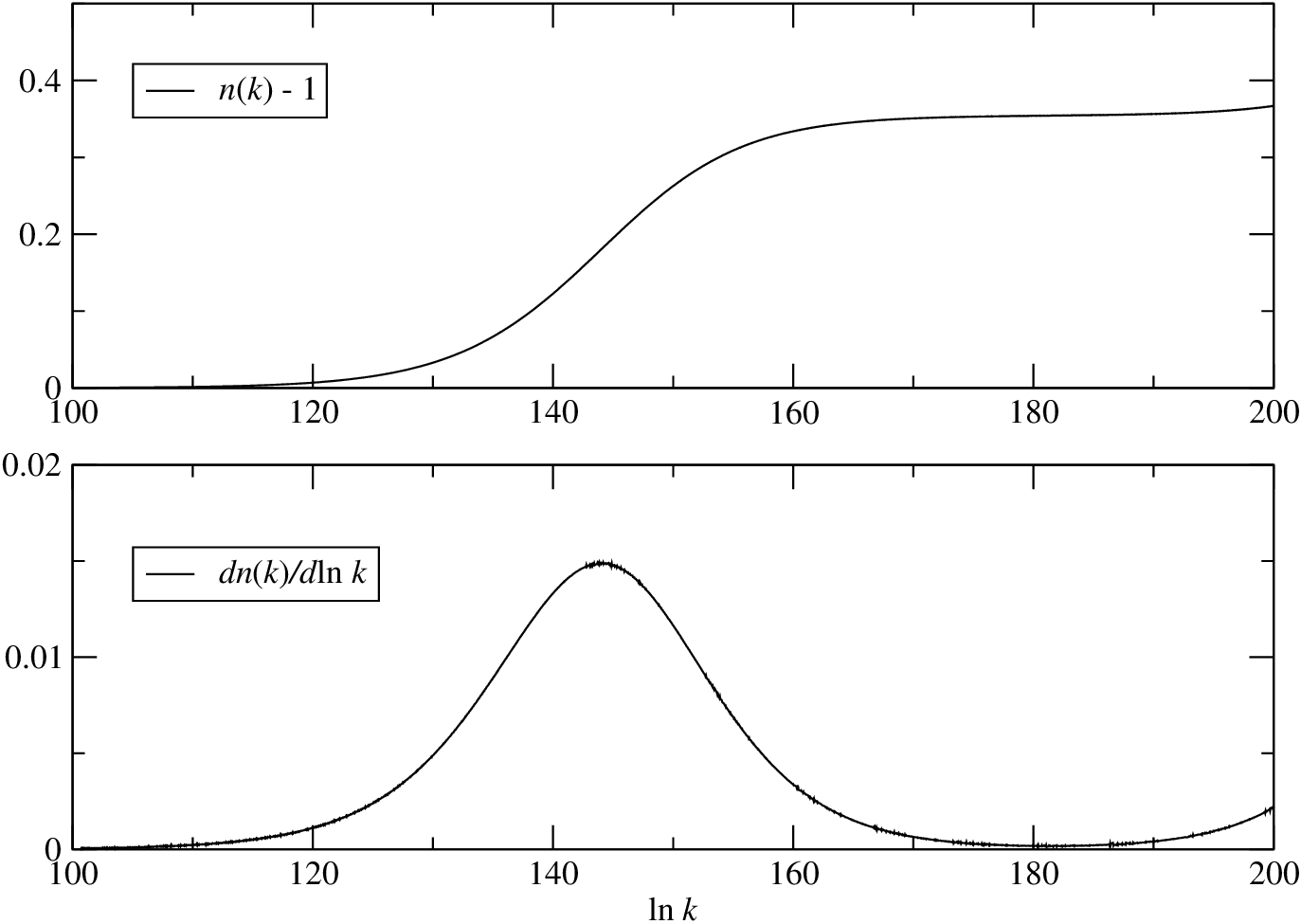}
\end{center}
\caption{Spectral index $n(k)$ and its running $dn(k)/d\ln k$ vs.
$\ln k$ ( $k$ in units of $m$ ) with initial conditions
$\phi(0)\approx 0$, $\dot{\phi}(0)=0$, and the parameters
$h=\frac{H}{m} =2.0$, $\lambda=10^{-14}$ in the case of $N=4$.}
\label{fig4}
\end{figure}

The correlation of fluctuations caused by the off-equilibrium
process is generic, because the evolution of fluctuations is
typically with respect to a time-dependent background. When
perturbations of scale $k$ cross the horizon at a time $t$, the
equal-time two point correlation function between $({\bf x},
t)$-$({\bf x'},t)$ will yield the correlation between two
space-scale modes $({\bf x}, k)$-$({\bf x'}, k)$ via the mapping
formula $k=a(t)H$ in Eq.(\ref{masspert}). The discrete wavelet
transform (DWT) is designed to do such space-scale [{\bf x}-{\bf
k}] decomposition~\cite{dau}.

In the formulation of DWT, there are two sets of spatially
localized bases given by the scaling functions $|{\bf
j,l}\rangle_s$, and the wavelet functions $|{\bf j,l}\rangle_{w}$;
both are characterized by the indices ${\bf j}$ and ${\bf l}$
where  ${\bf j}=0, 1, 2,...$ stands for a scale from ${\bf k_j}$
to ${\bf k_j} +\Delta {\bf k_j }$ in which ${\bf k_j} =2\pi2^{\bf
j}/L$ and $\Delta {\bf k_j} =2\pi2^{\bf j}/L$. The index $ {\bf
l}=0, 1,...,2^{\bf j}-1$ denotes the location of the spatial point
within $L{\bf l}/2^{\bf j} <{\bf x}_{\bf l} < L({\bf l}+1)/2^{\bf
j} $. These bases are complete, and they satisfy the orthogonal
relations ${}_s\langle { {\bf j}', {\bf l}'}|{ {\bf j},{\bf
l}}\rangle_s=\delta_{ {\bf j},{\bf j}'}\delta_{ {\bf l},{\bf
l}'}$, and ${}_w\langle { {\bf j}',{\bf l}'}|{ {\bf j},{\bf
l}}\rangle_w=\delta_{ {\bf j},{\bf j}'}\delta_{ {\bf l},{\bf
l}'}$. Then , the two-point correlation functions of the density
contrast $\delta $ can be rewritten in terms of the DWT bases as
\begin{eqnarray}
\langle \tilde {\delta}_{\bf j,l}\tilde{\delta}_{\bf j,l'}\rangle
 &=&  \int \frac{d^3k}{(2\pi)^3} \, \langle \delta( {\bf k},t) \,
 \delta(-{\bf k},t) \rangle \,
\hat{\psi}_{\bf j,l}({\bf k})\hat{\psi}^*_{\bf j,l'}({\bf k}) \, ,
\nonumber \\
\langle \delta_{\bf j,l}\delta_{\bf j,l'}\rangle
 &=&  \int \frac{d^3k}{(2\pi)^3} \, \langle \delta( {\bf k}, t)  \,
 \delta(-{\bf k},t) \rangle \,
\hat{\phi}_{\bf j,l}({\bf k})\hat{\phi}^*_{\bf j,l'}({\bf k}) \, ,
\nonumber
\\ \label{DWTcorrelation}
\end{eqnarray}
 where
$\hat{\psi}_{\bf j,l}({\bf k})=\langle {\bf k}|{\bf
j,l}\rangle_s$,and $\hat{\phi}_{\bf j,l}({\bf k})=\langle {\bf
k}|{\bf j,l}\rangle_w$ are the scaling functions and the wavelet
functions in the $k$-representation.  The time $t$ is taken to be
$t_j$ specified by the relation $2\pi 2^j/L=k=aH$. Since $a
=\exp(Ht) $, and  one has
\begin{equation}
t_j=\frac{1}{H}\left [j\ln 2 + \ln \left(\frac{2\pi}{LH}\right
)\right ] \, .
\label{tjrelation}
\end{equation}
Thus, Eq.(\ref{DWTcorrelation}) can be used to determine the
correlations between fluctuations at different spatial points
${\bf l}$ and ${\bf l'}$, both crossing out of the Hubble horizon
at the same time $t_j$ during the inflationary epoch.

\begin{figure}[t]
\begin{center}
\leavevmode \epsfxsize=2.0 in \epsffile{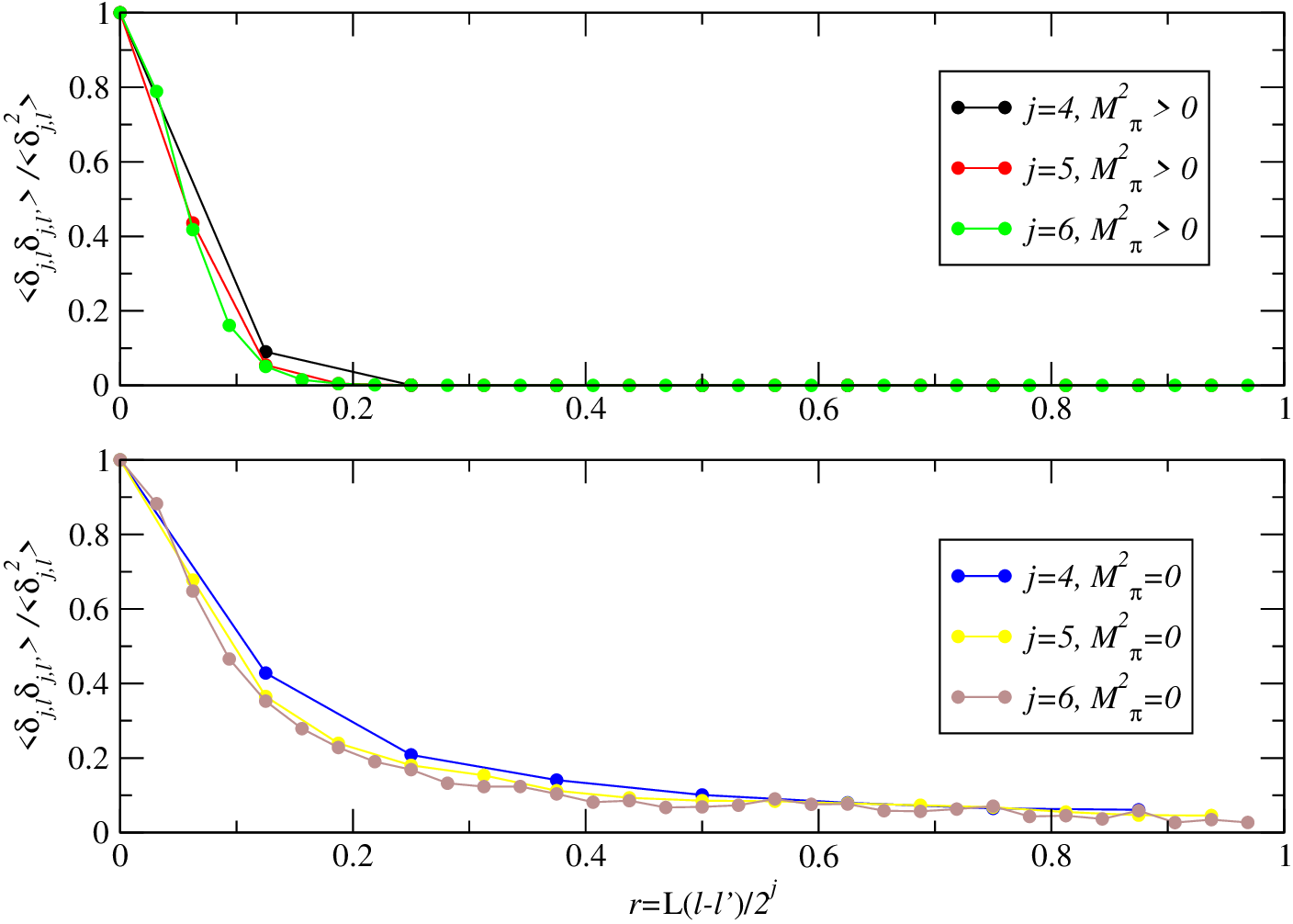} \caption{The
normalized mode-mode correlation functions of density fields in
phase space are plotted against the separation $r$ ( in units of
$m^{-1}$ ) of two perturbation modes with respect to various
scales $j$ under two different sets of initial conditions. The
parameter $L$ is taken to be $2m^{-1}$. } \label{fig7}
\end{center}
\end{figure}

As a numerical example, the normalized two-point space-scale
correlations of density perturbations defined as
$\langle\delta_{\bf j,l}\delta_{\bf j,l'}\rangle/
\langle\delta^2_{\bf j,l}\rangle$ under different initial
conditions are plotted in Fig.~\ref{fig7}, in which the parameters
are taken to be $H=2m$ and $L=2m^{-1}$. From
Eq.(\ref{tjrelation}), one has $t_j \simeq (1.63 - 2.33)m^{-1}$
for $j=4 - 6$  which corresponds to the number of e-foldings
$Ht_j\simeq 3.3 - 4.7$. Since the inflation under consideration
lasts from $t=0$ to about $ t= 80 m^{-1}$, the correlations in
Fig.~\ref{fig7} actually probe the inflaton dynamics at the
beginning of the inflation. With $M^2_{\chi}(0)=M^2_{\pi}(0)\simeq
m^2>0$, the space-scale correlation function approaches zero
drastically as the distance $r$ between the two modes increases;
while the correlation deviates from zero and lasts for a long
range for perturbations with $M^2_{\chi}=M^2_{\pi}=0$ initially.
It indicates that the  evolution of the correlation depends upon
the initial conditions of the inflation. From Eq.
(\ref{tjrelation} ), the $j$-dependence of the mode-mode
correlation can probe
 the evolution ($t$-dependence) of the
primordial perturbations. Thus,  one is capable of exploring the
physics of the very early universe by means of the mode-mode
correlations in phase space.

In summary, we study the off-equilibrium effects on the primordial
perturbations in the O(N) model. When the interaction or
self-interaction of inflaton becomes important, one must consider
the evolution of the inflationary background field and its
fluctuations with the back-reaction effects using a
self-consistent off-equilibrium formalism.  We find  two
observable remains left behind the off-equilibrium processes which
would serve as the  imprints to probe the epoch of inflation. The
first one is the running spectral index of primordial
perturbations. We find that the running spectral index depends
essentially on the rate of  the energy transfer from the
background field  to the inflaton fluctuations. The second remain
is the correlation  between phase space modes of the density
perturbations.  Under the influence of the self-interaction,
fluctuations created from the background field are no longer white
noises.  Moreover, since the evolution of the correlation depends
upon the initial conditions of the inflation, the mode-mode
correlation of density perturbations also provides a window to
study the dynamics of the self-interaction {\it as well as} the
onset of the inflation. Thus, we may expect that the non-trivial
mode-mode correlation in the phase space is detectable via a DWT
analysis on the CMB temperature map, or other observations on the
large scale structure relevant to the density
perturbations~\cite{pvf}.

DSL would like to thank Sung-Won Kim for organizing this wonderful
and very stimulating conference. This work was supported in part
by the NSC, Taiwan, R.O.C under the Grant NSC92-2112-M-001-029
(WLL), NSC92-2112-M-001-030 (YYC) and NSC91-2112-M-259-005 (DSL),
and by
 the NCTS (NSC
91-2119-M-007-004-), Taiwan, R.O.C..


\end{document}